\documentclass[superscriptaddress,twocolumn]{revtex4}
\usepackage{mathrsfs}
\usepackage{mathrsfs}
\usepackage{mathrsfs}
\usepackage{mathrsfs}
\usepackage{amssymb}
\usepackage[tbtags]{amsmath}
\usepackage{graphicx}
\usepackage{epsfig,graphicx,times}
\usepackage{color}
\usepackage{subfigure}
\usepackage{subeqnarray}
\usepackage{cases}

\begin{document}
\title{Enhancing the detection probability of single waveguided-photon by cavity technique}
\author{L. Y. Xie}
\affiliation{State Key Laboratory of Optoelectronic Materials
and Technologies, School of Physics and Engineering, Sun Yat-sen
University, Guangzhou 510275, China}

\author{L. F. Wei\footnote{E-mail: weilianfu@gmail.com}}

\affiliation{State Key Laboratory of Optoelectronic Materials
and Technologies, School of Physics and Engineering, Sun Yat-sen
University, Guangzhou 510275, China}
\affiliation{Quantum Optoelectronics Laboratory, Southwest Jiaotong University, Chengdu 610031, China}

\begin{abstract}
The resonant-cavity-enhanced (R\/C\/E) technique is an important
approach to increasing the detection efficiency (DE) of typical
free-space coupling photons. Here, we show that such a technique can
also be utilized to increase the detection probability (DP) of a
single waveguide-coupled photon. Based on a fully quantum mechanical
theory in real space, we exactly calculated the absorption probability
of a single photon for a two-level detector next to the waveguide. We
find that the DP of the waveguide photon for the detector in a
waveguide-coupled ring cavity is significantly higher than that for
the bare detector directly coupled to the photon. Physically, the DP
of the photon for the bare detector next to the waveguide is always limited by the finite transmission and reflection probabilities of the photon. The cavity technique is used to store the photon and thus increase its DP. The feasibility of the proposal with current integrated optical devices is then discussed.

PACS number(s):
32.80.Qk, 
42.50.Ct, 
42.79.Gn  
\end{abstract}

\maketitle
\section{Introduction}
Single-photon detection is a well-studied but nonetheless hot topic in
optical physics because it is related to various important
applications, specifically the realizations of the optical quantum
computation\cite{sci2007,RoMP2007}, quantum information\cite{natp20093} and precise
measurements~\cite{OE2014}. Photon detections
have been achieved by utilizing photomultiplier tubes~(P\/M\/T),
wherein the absorptions of the incident photons were converted into
detectable electronic signals. In semiconducting detectors,
specifically single-photon avalanche photon diodes
(S\/A\/P\/D)~\cite{mod2004}, the signals related to
photon-induced electron-hole pairs are probed. To decrease unwanted dark counts, low temperature detectors with quantum dots~\cite{nature2000} and superconductors~\cite{appl1998,JOMO2009} have also been developed.

\begin{figure}[!ht]
\centering%
\includegraphics[width=16pc]{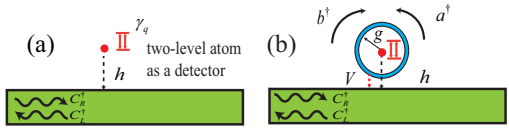}
\caption{\small Configurations of the detectors for single photons in the waveguide; (a) a bare two-level atom as a single-photon detector and (b) a two-level atom in a microring cavity as a single-photon detector. Here, the single-mode waveguide is denoted by the green channel, and the microring cavity is denoted by a blue circle.
The two-level atomic detector is denoted by a red dot.}
\label{fig_env2}
\vspace*{-5pt}
\end{figure}

Recently, to decrease the loss of photons and achieve convenient
optical device integrations
on a chip, integrated photonic
circuits~\cite{sci2008,LPRev2012} have been paid substantial
attention. In the waveguide, the photons are effectively confined in certain micro-structures and thus the relevant propagation losses can be minimized. Therefore, by integrating the waveguide and photonic devices simultaneously on a
chip~\cite{apl201097}, the efficiencies of the photonic emission, traveling, routing, detection, etc. can be significantly increased. Although the transport properties of single photons traveling along waveguide structures have been extensively investigated; see,
e.g., ~\cite{prl2007,pra20091,pra20092}, the research on the
detection of the waveguide photons with integrated photon detectors is
relatively
sparse~\cite{apl2011,Natc2012,apl2013,IEEE2015,IEEE201525,natc2015,pra2011,oe2013}.
Typically,
superconducting nanowire single photon detectors
(SNSPDs)~\cite{apl2011,Natc2012,apl2013,IEEE2015,IEEE201525,natc2015}and transition
edge sensors (TESs)~\cite{pra2011,oe2013} have been demonstrated for the
waveguide-coupled photon detections. Note that the $20\%$  DE (at telecommunication
wavelengths)~\cite{apl2011,Natc2012,IEEE2015}) of a single
photon for $GaAs$ SNSPD integrated waveguide-coupled detectors~\cite{IEEE2008} has been
experimentally demonstrated, which is substantially higher
than those for the free-space coupling counterparts (usually $\sim
10\%$~\cite{rofs2011}). A higher DE could be achieved by
further optimizing integrated waveguide-coupled detectors via further decreasing the coupling losses.

Given that the resonant cavity enhanced (R\/C\/E) technique has been extensively utilized to increase the DE of the usual free-space coupling photon detectors~\cite{jap1995}, it is expected that such a technique will also useful for waveguide-coupled photon detectors. The applied R\/C\/E devices can store the detected photons in the cavity, and thus, the relevant DE can be increased. Various resonant cavity geometries, including the Fabry-Perot resonant cavity~\cite{jap1995} and microring resonant cavity~\cite{ieeept2006}, have been successfully applied to increase the DEs of free-space-coupled photon detectors, {i.e., \it the photons propagate in free space until being absorbed by the detectors}.

In this paper, we first calculate the DP of the single photon confined in the waveguide
by a bare two-level atom. It is shown that such a probability is
quite low as a relatively low probability of the photon-induced atomic
excitation. To enhance the probability of atom excitation, we propose
an approach by using the RCE
technique, i.e., a resonant microring cavity
structure~~\cite{pra20092,pra1998,pra20075}(see also
Fig.~1(b)), to transport the traveling photon along
the waveguide into the cavity. Our discussion is based on a full
quantum theory~\cite{prl2007,pra20091,pra20092} for single
photons transporting in real space and interacting with a two level
detector in the resonant cavity. The detection of the single photon is
implemented by probing the excitation probability of the two-level
atomic detector coupled to the waveguide. Because the traveling waveguide photon is
converted into a standing wave that is stored in the cavity, the DP of the single photon can be increased. The feasibility of the proposal is also discussed.

\section{DP of a single photon in a waveguide by a bare two-level atomic detector}

Following Ref.~\cite{pra20091,pra20092,3564886}, the effective Hamiltonian describing a single photon propagating along a one-dimensional waveguide scattered by a single two-level atom can be expressed as
\begin{small}
\begin{eqnarray}
H_{1}/\hbar & = &-iv_{g}\int dx\{C_{R}^{\dagger}(x)\frac{\partial}{\partial x}C_{R}(x)-C_{L}^{\dagger}(x)\frac{\partial}{\partial x}C_{L}(x)\} \nonumber\\
 &+&\{\int dx\delta(x)\left[hC_{L}(x)\sigma^{\dagger}+hC_{R}(x)\sigma^{\dagger}\right]
 + h.c\}\nonumber\\
 &+&(\Omega_{e}-i\gamma_{q})\omega_{e}^{\dagger}\omega_{e}
 +\Omega_{g}\omega_{g}^{\dagger}\omega_{g}.\nonumber
\end{eqnarray}
\end{small}
Here, $C_{R/L}^{\dagger}(x)$ represents the bosonic creation operator at the position $x$ of the photon traveling in the right/left direction. $\Omega_{e}-\Omega_{g}\equiv\Omega$ is the atomic transition frequency, $\omega_{g/e}$ represents the ground/excited state frequency of the atom, and $\sigma^{\dagger}(\sigma)=\omega_{e}^{\dagger}\omega_{g}\,(\omega_{e}\omega_{g}^{\dagger})$ is
the creation (annihilation) ladder operator of the atom, $v_{g}$ is the group
velocity of the incident photon in the waveguide, $\gamma_q$ is the dissipation
rate of the excited atom, and $h$ (with units of $\sqrt{m}$ MHz to ensure consistent dimensions, more details can be seen in the Ref (~\cite{3564886}) is the efficient coupling strength between the atom and the
waveguide photon. Obviously, the first and second parts in the above Hamiltonian describe
the photon freely transporting along the waveguide and the bare atom, respectively.
The third part describes the interaction between the atom and the photon in the
waveguide.

The generic quantum state of the system can be expressed as
\begin{small}
\begin{eqnarray}
|\Phi_{1}(\tau)\rangle&=&\int dx\{\phi_{R}(x,\tau)C_{R}^{\dagger}(x)+\phi_{L}(x,\tau)C_{L}^{\dagger}(x)\}
|0,-\rangle\nonumber\\
&+&e_{q}(\tau)\sigma^{\dagger}|0,-\rangle,
\end{eqnarray}
\end{small}
which satisfies the Schr\"odinger equation
\begin{small}
\begin{eqnarray}
H_1|\Phi_1(\tau)\rangle=i\hbar\frac{\partial}{\partial \tau}|\Phi_1(\tau)\rangle.
\end{eqnarray}
\end{small}
Above, $ |0,->$ depicts the state of the system without any propagating photon in the waveguide, and the atom is in its ground state.
$\phi_{R/L}(x,\tau)$ is the single-photon wavefunction
transporting in the $R/L$ mode, and $e_{q}(\tau)$ the excitation amplitude of the atom.
The system wave function can be rewritten as $|\Phi(\tau)\rangle=e^{-i\epsilon \tau}|\epsilon\rangle$, with $\epsilon$ being the eigenfrequency, and
\begin{small}
\begin{eqnarray}
|\epsilon\rangle&=&\int dx\left[\phi_{R}(x)C_{R}^{\dagger}(x)
 +\phi_{L}(x)C_{L}^{\dagger}(x)\right]|0,-\rangle\\ \nonumber
 &+&e_{q}\sigma^{\dagger}|0,->.
\end{eqnarray}
\end{small}
The coefficients in Eq.~(4) are determined by
\begin{small}
\begin{eqnarray}
-iv_{g}\frac{\partial}{\partial x}\phi_{R}(x)+\delta(x)he_{q}+\Omega_{g}\phi_{R}(x) &=&\epsilon\phi_{R}(x),\nonumber\\
+iv_{g}\frac{\partial}{\partial x}\phi_{L}(x)+\delta(x)he_{q}+\Omega_{g}\phi_{L}(x)  &=&\epsilon\phi_{L}(x),\\
(\Omega_{e}-i\gamma_{q})e_{q}+h^{*}\phi_{R}(0)+h^{*}\phi_{L}(0)&=&\epsilon e_{q}\nonumber,
\end{eqnarray}
\end{small}
with $\epsilon=\omega+\Omega_{g}$ and $\omega=kv_{g}$. Introducing the
Heaviside step function $\theta(x)$,
we have  $\phi_{R}(x)=e^{ikx}[\theta(-x)+t\theta(x)]$ and $\phi_{L}(x)=re^{-ikx}\theta(-x)$,
with $t$ and $r$ being the transmission and reflection amplitudes of the photon,
respectively. Consequently, Eq.~(5) can be further simplified as
\begin{small}
\begin{eqnarray}
-iv_{g}(t-1)+he_{q}&=&0,\nonumber\\
+iv_{g}(-r)+he_{q}&=&0,\\
(\Omega-i\gamma_q)e_{q}+h^{*}\frac{1+t}{2}+h^{*}  \frac{r}{2}&=&\omega e_{q}\nonumber
\end{eqnarray}
\end{small}
To simplify the calculation, we have set the coupling strength between the waveguide and atom is real, i.e. $h=h^*=|h|$. The solution of the above equation reads
\begin{small}
\begin{eqnarray}
t=  \frac{(i\gamma_q+\omega-\Omega)}{(i\gamma_q+\omega-\Omega)
+i\Gamma_{1}},\nonumber\\
\nonumber
r=  \frac{-i\Gamma_1}{(i\gamma_q+\omega-\Omega)+i\Gamma_{1}},\\
\nonumber
e_{q}= \frac{h}{(i\gamma_q+\omega-\Omega)+i\Gamma_{1}},
\end{eqnarray}
\end{small}
with $\Gamma_{1}=h^2/v_{g}$. Generically, the normalized DP $\eta_b$ of the photon reads
\begin{small}
\begin{eqnarray}
\eta_b & = & \frac{|e_{q}|^{2}}{|t|^{2}+|r|^{2}+|e_{q}|^{2}}.
\end{eqnarray}
\end{small}
\begin{figure}[!ht]
\centering%
\includegraphics[width=20pc]{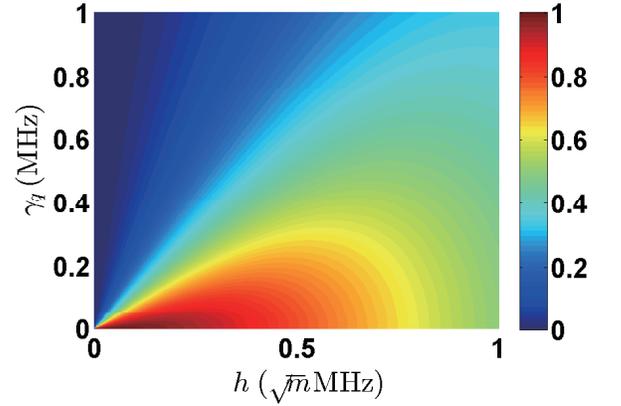}
\caption{DP $\eta_b$ of a single confined photon by a bare two-level detector versus the atomic dissipation rate $\gamma_q$ and the resonant (i.e., $\omega-\Omega =0$) coupling strength $h$ between the detector and photon. Here, $v_g=1$.}
\label{fig_env2}\vspace*{-6pt}
\end{figure}

\begin{figure}[!ht]
\centering%
\includegraphics[width=20pc]{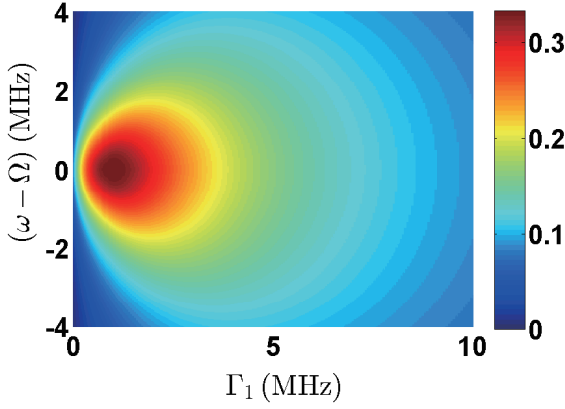}
\caption{DP of the single confined photon in the waveguide by the atomic detector with a defined dissipation $\gamma_q/2\pi=0.16MHz$. The maximum of the DP is $\eta_b=33.22\%$ for the resonant photon.}
\label{fig_env2}\vspace*{-6pt}
\end{figure}

Fig.~2 shows how the excitation probability $\eta_b$ (i.e., the DP of
the confined photon by a bare two-level detector) varies with the
dissipation $\gamma_q$ of the excited atomic state and the
photon-detector resonant coupling strength $h$. For a defined atomic
dissipation rate $\gamma_q$, the value of $\eta_b$ is found to exhibit
a maximal value when varying the photon-detector interacting
strength. In addition, the peak value of $\eta_b$ increases
monotonously with decreasing atomic dissipation. The desired
photon detection is usually achieved by probing the leakage of the
population from the excited atomic state. This implies that the
dissipation of the detector atom should be sufficiently large;
otherwise, its leakage cannot be measured. Moreover, when
$\gamma_q\sim 0$, the resonant photon is fully reflected by the atomic detector, i.e., $|r|^2\sim 1$ and $|t|^2\sim 0$, even though the atom can still
be temporally excited. In this case, the photon cannot be detected by the detector.

For a typical dissipation of the atomic detector, e.g., $\gamma_q/2\pi=0.16$ MHz~\cite{pra20075}, Fig.~3 shows the DP of the single confined photon in the waveguide versus the effective atom-photon detuning $\Delta\equiv\omega-\Omega$ and the
effective atom-photon coupling strength $\Gamma_1=|h|^2/v_g$.
It is seen that, for the fixed dissipation $\gamma_q/2\pi=0.16$ MHz
of the atomic detector, the DP of the photon exhibits a maximal value
for certain parameters and varies symmetrically with the frequency of
the incident light. For a
defined $\Gamma_1$, the DP of the resonant photon is
always higher than that of the off-resonant photon. Furthermore, for
a defined atom-photon detuning, the DP of the photon exhibits a peak
value, corresponding to an optimal effective atom-photon coupling. The maximum DP of the resonant photon by the present bare detector is approximately $\eta_b=33.22\%$. This is low for desired quantum information processing on chip and should be increased.

\section{Cavity-enhanced DP of a single resonant photon in a waveguide}

It is well known that the microcavity technique has matured in recent
decades. The RCE devices benefit from
the wavelength selectivity and increased optics-matter
interactions. As a consequence, RCE photodetectors can be thinner and
smaller, thereby increasing the sensitivity of the photon response and thus the quantum efficiency of the photon detection. With such a technique, light with off-resonant wavelengths is rejected. In this section, we show that such a technique can also be utilized to increase the excitation probability of a single two-level atomic detector for waveguide photons.

The basic idea of the RCE technique is to convert a photon
traveling along the waveguide into a standing wave in a cavity. We assume that the bare atomic detector is placed in a microring
cavity with two degenerated whispering-gallery modes (WGMs), namely, $a^\dagger$
and $b^\dagger$ modes. Now, the effective Hamiltonian of the system reads
\begin{small}
\begin{eqnarray}
H_{2}/\hbar & = &-iv_{g}\int dx
\{C_{R}^{\dagger}(x)\frac{\partial}{\partial x}C_{R}(x)-C_{L}^{\dagger}(x)\frac{\partial}{\partial x}C_{L}(x)\}\nonumber \\
 &  & +(\Omega_{e}-i\gamma_q)\omega_{e}^{\dagger}\omega+\Omega_{g}\omega_{g}^{\dagger}\omega_{g}\nonumber\\
 &  &
 +(\omega_{c}-i\gamma_{c})a^{\dagger}a+(\omega_{c}-i\gamma_{c})b^{\dagger}b\nonumber\\
 &  & +\{\int dx\delta(x)h[C_{L}(x)\sigma^{\dagger}+C_{R}(x)\sigma^{\dagger}]\nonumber\\
 &  & +\int dx\delta(x)[V_{a}C_{R}(x)a^{\dagger}+V_{b}C_{L}(x)b^{\dagger}]\nonumber\\
 &  & +[g_{a}a+g_{b}b]\sigma^{\dagger}+h.c\}.
 \end{eqnarray}%
 \end{small}
Here, $a^{\dagger}$ and $ b^{\dagger}$ are the Bosonic operators
describing the anticlockwise and clockwise WGMs in the microring
cavity with the same frequency $\omega_{c}$ and dissipation rate
$\gamma_c$, respectively. The sixth and seventh terms in the Hamiltonian (9) are the interactions between the waveguide-cavity modes, and $V_{a/b}$ and $g_{a/b}$ describe the coupling strength between the $a/b$ mode and the photon in the waveguide and the atom detector, respectively.
Note that the efficient coupling strength $V$ has the same dimensions as $h$, and the unit
of $g$ is that of the frequency. The generic wave function of the present system takes the form%
\begin{small}
\begin{eqnarray}
|\Phi_{2}(\tau)\rangle& = &\int dx\{\phi_{R}(x,\tau)C_{R}^{\dagger}(x)+\phi_{L}(x,\tau)C_{L}^{\dagger}(x)\}|0,0,-\rangle \nonumber \\
 &  &  +e_{a}(\tau)a^{\dagger}|0,0,-\rangle+e_{b}(\tau)b^{\dagger}|0,0,-\rangle\nonumber\\
 &  &+e_{q}(\tau)\sigma^{\dagger}|0,0,-\rangle,
\end{eqnarray}%
\end{small}
where $|0,0,-\rangle$ represents the ground state of the system (i.e., there is no
propagating photon in the waveguide, and the cavity and the atom are not excited).
With the same approach used in Sec.~II, for a single photon input
along the waveguide from the left direction, the excitation
probabilities of the atomic detector and the WG modes as well as the
reflection and transmission amplitude of
the incident photon are determined as follows:
\begin{small}
\begin{eqnarray}{}
-iv_{g}(t-1)+V_{a}e_{a}+he_{q}=0,\nonumber \\
+iv_{g}(-r)+V_{b}e_{b}+he_{q} =0,\nonumber\\
(\omega_{c}-i\gamma_{c})e_{a}+V_{a}^{*}\frac{1+t}{2}+g_{a}e_{q}=\omega e_{a},\\
(\omega_{c}-i\gamma_{c})e_{b}+V_{b}^{*}\frac{r}{2}+g_{b}e_{q} =\omega e_{b}\nonumber\\
({\Omega}-i\gamma_{q})e_{q}+h^{*}\frac{1+t}{2}+h^{*}\frac{r}{2}+g_{a}^{*}e_{a}+g_{b}^{*}e_{b} =\omega e_{q}\nonumber.
\end{eqnarray}
\end{small}

Solving the above equation, we have
\begin{small}
\begin{eqnarray}{}
e_{q}&=&\frac{-2i\Lambda}{2i(2g^2-AB)+2\Theta+4hgV},\nonumber\\
e_{b}&=&\frac{-(2ig+hV)\Lambda}{(A+i\Gamma_{2})[2i(2g^2-AB)+2\Theta
+4hgV]},\nonumber\\
e_{a}&=&\frac{V[2i\Upsilon+2\Theta+4hgV]-(2ig+Vh)\Lambda}
{(A+i\Gamma_2)[2i\Upsilon+2\Theta+4hgV]},\\
r&= &\frac{-2\Lambda^2}{(A+i\Gamma_{2})
[2i\Upsilon+2\Theta+4hgV]},\nonumber\\
t&=&\frac{(A-i\Gamma_2)[2i(2g^2-AB)+2\Theta+4hgV]-2\Lambda^2}
{(A+i\Gamma_{2})[2i\Upsilon+2\Theta+4hgV]}\nonumber,
\end{eqnarray}
\end{small}
with $\Lambda=Ah+gV,\,A=(\omega-\omega_{c}+i\gamma_{c})$, $\Upsilon=(2g^2-AB),\,\Theta=A\Gamma_1+B\Gamma_2,\,B=(\omega-\Omega+i\gamma_q)$, and $\Gamma_{1}=h^2/v_{g}, \Gamma_{2}=V^2/2v_{g}$. We have assumed $V_{a}=V_{b}=V=V^*$, $g_{a}=g_{b}=g=g^*$ and $v_g=1$ for simplicity. Consequently, the normalized DP of the photon in the waveguide by the atom detector in the cavity reads

\begin{equation}
\eta_c=\frac{|e_{q}|^{2}}{|t|^{2}+|r|^{2}+|e_{q}|^{2}+|e_{a}|^{2}+|e_{b}|^{2}}.
\end{equation}

For the optimized atom-cavity coupling strength $g/2\pi=0.29$MHz, Fig.~4 shows how the DP of the resonant photon depends on the coupling strength
$V$ (between the photon in the waveguide and the cavity) and $h$
(between the atom and the photon in the waveguide) for typical dissipation rates for the atom and cavity~\cite{pra20075}: $\gamma_q/2\pi=0.16$ MHz and $\gamma_{c}/2\pi=0.76$ MHz.
Specifically, one can observe the following:

i) For a fixed coupling strength $h$,
the DP obtains a peak value for certain selected values of $V$. However, for $V/2\pi\geq0.6$ $\sqrt{m}$ MHz, the DP monotonously decreases.

ii) When $V\sim0$, the photon can still be  directly detected by the atom detector, although the probability is relatively low, e.g., the maximal value is $\eta_c\sim 0.2$.

iii) Interestingly, there exists a cavity-enhanced ($h, V$)-regime (i.e., the yellow
regime with $\eta_c\geq 0.34$), wherein the DP of the photon in the waveguide is
significantly increased compared to that for the bare atom detector.
\begin{figure}[!ht]
\centering
\includegraphics[width=20pc]{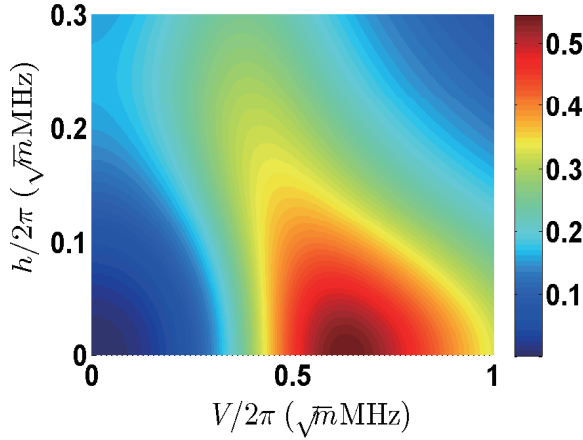}
\caption{DP $\eta_c$ of the photon for a two-level atomic detector in a microring cavity versus the coupling strength
$V$ (between the photon in the waveguide and the cavity) and $h$
(between the atom and the photon in the waveguide). The other
parameters are  $\gamma_q/2\pi=0.16$ MHz, $\gamma_{c}/2\pi=0.76$
MHz~\cite{pra20075}, and $g/2\pi=0.29$ MHz when we properly varied to achieve the maximum DP.}
\label{fig_env2}\vspace*{-6pt}
\end{figure}

We now analyze the maximum value of $\eta_c$, which is found (in Fig.~4)
to be $\eta_c=54.39\%$ for $h/2\pi=0$ and $V/2\pi=0.61$$\sqrt{m}$ MHz.
This corresponds to the very weak atom-waveguide coupling, i.e., $h\sim 0$, and the phase- and magnitude-matching conditions
\begin{small}
\begin{eqnarray}
 \Gamma_2Vgh=0,
\end{eqnarray}
\end{small}
and
\begin{small}
\begin{eqnarray}
(\gamma_c-\Gamma_2)[2g^2+\gamma_q(\gamma_c+\Gamma_2)+\gamma_c\Gamma_1]+g^2V^2-h^2\gamma_c^2=0,
\end{eqnarray}
\end{small}%
respectively, are satisfied. As a consequence, the optimized coupling strength $g$
(between the atom and the cavity) should be designed as $g=\sqrt{(\Gamma_{2}^{2}-\gamma_c^2)\gamma_q}/\sqrt{2\gamma_c}\approx 1.82$ MHz, and thus, $g/2\pi\approx 0.29$ MHz.
Physically, the matching conditions demonstrated above imply that the photon in the
waveguide is maximally (with $82\%$ probability) converted in the cavity. Thus,
the DP of the photon for the atom detector in the cavity is maximal.
\begin{figure}[!ht]
\centering
\includegraphics[width=20pc]{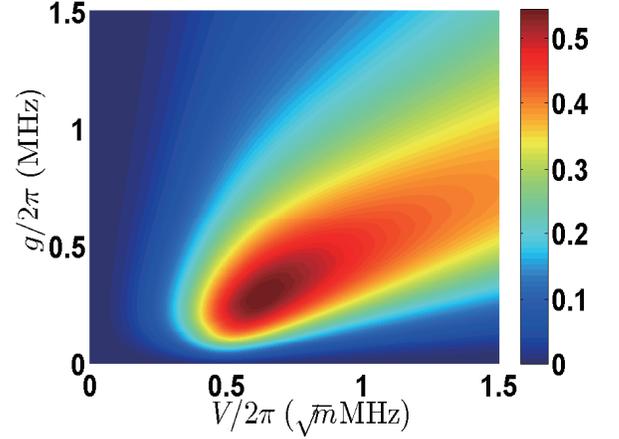}
\caption{
DP $\eta_c$ of the photon for a two-level atom detector in a microring cavity versus the coupling strength
$V$ (between the photon in the waveguide and the cavity) and $g$
(between the atom and cavity). The parameters are $\gamma_q/2\pi=0.16$ MHz, $\gamma_{c}/2\pi=0.76$ MHz, and $h=0$.}
\label{fig_env2}\vspace*{-6pt}
\end{figure}

This can be verified further using the numerical results shown in Fig.~5, which shows how the parameters $g$ and $V$ influence the DP of the photon. Here, the interaction between the atom and the photon in the waveguide is negligible due to the screen of the cavity.
Note that at the maximal DP point, the reflection probability $|r|^2$
of the photon in the waveguide remains non-zero (although $|t|^2\sim
0$) due to the existence of the atom detector in the cavity.

Fig.~6 shows that at the maximal DP point (with the optimized atom-cavity coupling strength $g=1.82$ MHz), the reflection probability of the photon is $|r|^2\approx 18.00\%$; the excitation probability of the $a$- and $b$ modes in the cavity are $|e_a|^2\approx26.81\%$ and $|e_b|^2\simeq 1.65\%$, respectively.

\begin{figure}[!ht]
\centering%
\includegraphics[width=20pc]{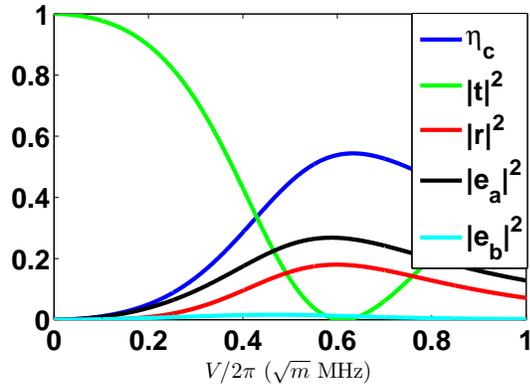}
\caption{DP and transport parameters of the photon in the waveguide, e.g., the detection, reflection and transmission probabilities etc. versus the coupling strength $V$. Here, the dissipation rates of the cavity and atom are set to $\gamma_q/2\pi=0.16$ MHz and $\gamma_{c}/2\pi=0.76$ MHz, and the coupling strength of the atom and the photon in the waveguide is $h=0$.}
\label{fig_env2}\vspace*{-6pt}
\end{figure}

\section{Conclusions}

In summary, we propose an approach based on a full quantum mechanic
theory in real space to increase the DP of a single photon
transporting along a one-dimensional waveguide. For a bare detector,
we found that the DP is related to the dissipation rate of the atom
and the atom-photon coupling strength. Typically, for a definite
atomic dissipation, e.g., $\gamma_q=0.16$ MHz, the maximum DP of the
resonant photon in the waveguide for the bare detector is
$33.22\%$. However, if the atom detector is placed adjacent to a
microring cavity, then the traveling photon can be transported into
the cavity and stored as standing wave modes. This increases the DP of
the photon. Indeed, our analytical and numerical results show that the
DP of the photon in a waveguide for the detector in a cavity can be
significantly increased, and the maximal value of this probability can
reach $54.39\%$. This again suggests that the RCE technique, used
successfully to increase the detection efficiency of the photons in
free space,
can also be utilized to enhance the DP of a single photon in waveguide structures.
In principle, by integrating a series of cavity as the reflectors~\cite{natc2015} the ideal detection (i.e., its DP approaches to $100\%$) of the waveguide photons is feasible.

Hopefully, our proposal is feasible and can be directly
applied to current integrated optoelectronics for optical quantum
information processing on chips. Integrating the waveguide device and
the photon detector on a chip is realizable with current integrated
photonic techniques~\cite{apl2011,Natc2012}; the quantum efficiency of
a waveguide nanowire superconducting single-photon detector has been
increased to $20\%$~\cite{IEEE2015}. In addition, in
Ref.~\cite{Jpd2009}, a system consisting of quantum dots (artificial
atoms could be used as photon detectors) in a cylindrical glass
waveguide were fabricated. Furthermore, WGMs have been observed in a
GaN-based microdisks~\cite{natp2007}. Therefore, with current
integrated optical techniques, the increased DP of the photon in a waveguide proposed in this paper should be feasible.

It is worth emphasizing that the physical mechanism presented here to
detect the waveguide photon is different from those in the
SSPDs~\cite{Natc2012}. Cooper pairs in superconductors are the
collective elementary excitations of electrons and are not
local. Photon detection by the SSPDs is implemented by probing the
disappearance of these elementary excitations by amplifying the
absorption effects of the photon. In the present model, a single
photon is detected by probing the induced excitation of a single
two-level atom. Therefore, in principle, the previously demonstrated
SSPDs should not be modeled simply as a single two-level
detector. However, whether a series of two-level detectors can be
integrated to describe the SSPDs remains an open question and requires
further investigation.

\section{Acknowledgments}
This work was supported in part by the National Fundamental Research Program of China through Grant No. 2010CB92304 and the National Science Foundation grant Nos. 11174373, 61301031, U1330201.

\end{document}